\documentclass[pre, showpacs]{revtex4}
\usepackage{graphicx,amsmath}
\def\beq{\begin{equation}}
\def\eeq{\end{equation}}
\def\bfg{\begin{figure}}
\def\efg{\end{figure}}

\begin{document}

\title{Subwavelength fractional Talbot effect in layered heterostructures of composite metamaterials}
\author{Simin Feng,  Klaus Halterman,  and Pamela L. Overfelt}
\affiliation{Physics and Computational Sciences  \\  Naval Air Warfare Center, China Lake, CA 93555}

\begin{abstract}
We demonstrate that under certain conditions, fractional Talbot revivals can occur in heterostructures of composite metamaterials, such as multilayer positive and negative index media, metallodielectric stacks, and one-dimensional dielectric photonic crystals.  Most importantly, without using the paraxial approximation we obtain Talbot images for the feature sizes of transverse patterns smaller than the illumination wavelength.  A general expression for the Talbot distance in such structures is derived, and the conditions favorable for observing Talbot effects in layered heterostructures is discussed.
\end{abstract}


\maketitle

\parindent.3in
The Talbot effect, or the repeated self-imaging of periodic patterns, is one of the most basic phenomena in optics.  As a consequence of Fresnel diffraction, periodic patterns can reappear (also called a revival image), upon propagation, at integer multiples of the so-called Talbot distance, $Z_T=2a^2/\lambda$, where $a$ is the spatial period of the pattern, and $\lambda$ is the wavelength of the incident light.  In addition to full revivals, fractional revivals with modified periods occur at distances that are rational fractions of the Talbot distance, i.e., $z/Z_T=p/q$, where $p$ and $q$ are coprime integers.  This remarkable self-imaging phenomenon has also been demonstrated in temporal\cite{Yeazell}-\cite{Deng} and spectral\cite{Wang1} domains in many areas of physics from classical optics to quantum matter waves, such as gradient-index lenses\cite{Flores}, waveguide arrays\cite{Iwanow}, atomic and molecular wave packets\cite{Yeazell,Vrakking}, and Bose-Einstein condensates\cite{Deng}.  Recent investigations have revealed that the Talbot effect is far more than a mere optical curiosity.  It is deeply connected to classical number theory and the intricate structure of physics\cite{Berry}.  Talbot effects have possible applications in optical free-space interconnects\cite{Leger}, integrated optics, and integer factorization schemes\cite{Clauser} in optical computing.  In parallel, several new types of composite metamaterials with sophisticated electromagnetic properties have been developed\cite{Shelby}-\cite{Klaus}.  These materials may lay the foundation for futuristic integrated optics and electronics.  The ability to control light in certain metamaterials has thus both scientific and practical importance.  It is therefore interesting to explore Talbot effects in composite metamaterials and potential applications involving the fractional Talbot effects, e.g., imaging in material fabrication to increase the spatial frequencies of periodic patterns.  One way to control wave diffraction is to construct multilayer photonic structures.  Although 1D periodic structures have been studied profusely, such systems still continue to reveal some interesting results\cite{Fink}-\cite{Feng1}.  Previous works have shown that photonic bands can not only control the transmission frequencies, but also deeply affect wave diffraction\cite{Longhi}-\cite{Manela}, leading to nondiffracting beams\cite{Manela} and high-resolution imaging\cite{Feng1}.  Typically Talbot images have been demonstrated in homogeneous media with feature sizes larger than the wavelength.  The higher order diffraction destroys the Talbot images when the feature sizes smaller than the wavelength.  In this letter, we show that through photonic engineering of wave diffraction, the fractional Talbot effect can be observed for subwavelength features of transverse patterns in inhomogeneous media, such as multilayer positive and negative index (PNI) materials, metallodielectric (MD) nanofilms, and low-dimensional double-dielectric (DD) photonic crystals.  A general expression for the Talbot distance in layered heterostructures is derived and confirmed numerically.

In any homogeneous medium, the monochromatic electromagnetic field vectors and the corresponding wavevector always form a right-handed set $(\bf E, B, k)$ independent of the constitutive relations of the materials.  Therefore, the continuity of the tangential field components $\bf E_{\parallel}$ and $\bf H_{\parallel}$ at the interface leads to the reversed sign of the normal component ($\bf k_{\perp}$) of the wavevector when the light is incident from a right-handed medium (RHM, $\epsilon>0$, $\mu>0$) to a left-handed medium (LHM, $\epsilon<0$, $\mu<0$), and vice versa, while the continuity of the tangential $\bf E_{\parallel}$ and the normal $\bf B_{\perp}$ leads to the continuity of the tangential component ($\bf k_{\parallel}$) of the wavevector at the interface.

In modeling heterostructures, we assume the permittivity and permeability to be constant for the positive index material and frequency-dependent for the negative index material.  The metal is assumed to have constant permittivity and frequency-dependent permeability.  The frequency-dependent permittivity and permeability are given by the Drude model:
\beq
\label{eps}
\epsilon_2 = 1 - {\omega_e^2\over\omega^2+i\omega\gamma_e},  \hspace{.5in}
\mu_2 = 1 - {\omega_m^2\over\omega^2+i\omega\gamma_m},
\eeq
where $\omega_e$ and $\omega_m$ are, respectively, the effective electric and magnetic plasma frequencies, and $\gamma_e$ and $\gamma_m$ are the corresponding damping factors.  The layer thicknesses are $d_1$ and $d_2$, respectively, for the positive and negative index materials.  The spatial period is $d=d_1+d_2$, thus, $\epsilon(z+d)=\epsilon(z)$ and $\mu(z+d)=\mu(z)$.  In our layered heterostructures, the index 1 refers to the positive index material while the index 2 refers to the negative index material, metal, or dielectric with higher refractive index.  We consider TM modes ($H_z=0$) and solve for the magnetic field out of convenience.  A similar result can be carried out for TE modes.  The vector wave equation for a monochromatic field ${\bf H=H(r)}\exp(-i\omega t)$ is $\nabla\times\{\epsilon^{-1}\nabla\times{\bf H}\}=(\omega/c)^2\mu{\bf H}$.  Thus, the equation for the $x$ and $y$ magnetic field components becomes\cite{Longhi}
\beq
\label{Hxy}
-\epsilon{d\over dz}\left( {1\over\epsilon} {d\over dz}H_{x,\,y}\right) + k_\perp^2 H_{x,\,y}
= \left({\omega\over c}\right)^2 \epsilon\mu\, H_{x,\,y}
\eeq
where $k_\perp^2=k_x^2+k_y^2$ is the transverse wave number and $k_xH_x+k_yH_y=0$.  The eigenmodes of the periodic structure are Bloch waves and the dispersion relation is given by\cite{Feng2}
\beq
\label{disp}
\cos(\beta d) = \cosh(\alpha_1d_1)\cosh(\alpha_2d_2) + 
{\alpha_1^2\epsilon_2^2 + \alpha_2^2\epsilon_1^2 \over 2\alpha_1\alpha_2\epsilon_1\epsilon_2}
\sinh(\alpha_1d_1)\sinh(\alpha_2d_2) \ ,
\eeq
where $\alpha_i^2=k_\perp^2-(\omega/c)^2\epsilon_i\mu_i,(i=1,2)$ and the $\beta$ is the Bloch wave number.  The existence of Bloch modes requires that $\bigl|\cos(\beta d)\bigr|\le1$.  It is well known that when this condition holds, the Bloch modes represent propagating waves.  Moreover, all-evanescent modes ($\alpha_i^2>0$ for $i=1$ and 2) can also exist\cite{Feng2}.  In such scenarios, the Bloch modes represent the transmission of coupled evanescent waves.  Equation~(\ref{disp}) also represents the diffraction relation in the multilayer medium from which the diffraction curve, $\beta$ versus $k_\perp$, can be derived.  Thus, wave diffraction in the multilayer structure can be constructed from a superposition of Bloch modes:
\beq
\label{H1}
H_{x,\,y}({\bf r}) = \int dk_x dk_y {\widetilde H_{x,\,y}(k_x,k_y)} u_\beta^*(0) u_\beta(z)\exp\left(ik_xx+ik_yy +i\beta z\right),
\eeq
where $\widetilde H_{x,\,y}(k_x,k_y)$ is the spectrum at the $z=0$ plane and $|u_\beta(0)|^2=1$.  The superposition in Eq.~(\ref{H1}) is limited to a single photonic band.  Since the Talbot distance ($Z_T$) is much larger than the period ($d$) of the multilayer structure, one can take the image at the distance of an integer multiple of the period closest to the Talbot distance, i.e. $Z_T=md+\delta\approx md$, where $m$ is a large integer and $\delta\ll Z_T$.  Using the fact that $u_\beta(md)=u_\beta(0)$, Eq.~(\ref{H1}) can be simplified as\cite{Note}
\beq
\label{H2}
H_{x,\,y}({\bf r}) = \int dk_x dk_y {\widetilde H_{x,\,y}(k_x,k_y)} \exp\left(ik_xx+ik_yy +i\beta z\right).
\eeq
To produce the fractional Talbot effects from Eq.~(\ref{H2}) requires a quadratic diffraction upon propagation.  This can be obtained by properly choosing parameters that satisfy Eq.~(\ref{disp}).

To effectively choose the system parameters and for comparison purposes, we also evaluate the dispersion relation (Eq.~(\ref{disp})) within the paraxial approximation (whereby $k_\bot\ll\beta_0$, where $\beta_0$ is the center Bloch wave number $\beta_0\equiv\beta(k_\perp=0$)).  We thus take the Taylor expansion of Eq.~(\ref{disp}) at $\beta=\beta_0$:
\beq
\label{beta}
\beta = \beta_0 - {1\over2d\sin(\beta_0d)} \left\{ {\partial^2Y\over\partial k_x^2} k_x^2 + {\partial^2Y\over\partial k_y^2} k_y^2\right\}
+ \vartheta(k_\perp^4),
\eeq
where $Y$ is the right hand side of Eq.~(\ref{disp}) and $d$ is the period of the structure.  In the above expansion, the odd derivatives vanish since $Y$ is an even function.  The fourth derivatives arising in the higher order terms are given by
\begin{eqnarray}
\label{beta4}
{\partial^4\beta\over\partial k_j^4} \biggl|_{k_\bot=0} &=& -{3\cos(\beta_0d)\over d\sin^3(\beta_0d)} \left({\partial^2Y\over\partial k_j^2}\right)^2
- {1\over d\sin(\beta_0d)} {\partial^4Y\over\partial k_j^4}, \hskip.3in  j = x,y,  \\
\label{beta44}
{\partial^4\beta\over\partial k_x^2 \partial k_y^2} \biggl|_{k_\bot=0} &=& 
-{\cos(\beta_0d)\over d\sin^3(\beta_0d)} {\partial^2Y\over\partial k_x^2} {\partial^2Y\over\partial k_y^2}
-{1\over d\sin(\beta_0d)} {\partial^4Y\over\partial k_x^2\partial k_y^2} \ .
\end{eqnarray}

\noindent
Near the band edges, $\sin(\beta_0d)\approx0$, and subsequently the first terms in Eq.~(\ref{beta4}) and Eq.~(\ref{beta44}) grow rapidly.  Thus, quadratic diffraction is achievable around the middle of the transmission bands with a reasonable spatial bandwidth.  Figure~\ref{FengFig1} shows the diffraction curves obtained from the exact formula Eq.~(\ref{disp}) and from the quadratic approximation, Eq.~(\ref{beta}), for the MD (Fig.~\ref{FengFig1}a) and DD (Fig.~\ref{FengFig1}b) stacks.  In Fig.~\ref{FengFig1}, the diffraction curve in the two heterostructures is nearly quadratic.  For comparison purposes, the diffraction curve of free space and its quadratic approximation are shown.  All the curves are terminated at the cut-off frequency of the corresponding medium (beyond that the spatial components cannot be transmitted).  In the PNI stack, the phase compensation effect of the negative index material leads to diffraction compensation.  Hence, quadratic diffraction can also appear near band edges when approaching the non-diffraction limit.  Figure~\ref{FengFig2} shows the diffraction curves near the middle of the band (Fig.~\ref{FengFig2}a) and near the band edge (Fig.~\ref{FengFig2}b) in the PNI stack.  The coincidence of the exact diffraction with the quadratic curve indicates quadratic diffraction inside the PNI stack.  Also the diffraction in the PNI stack is much less than that of free space.  Notice that in Figs.~\ref{FengFig1} and \ref{FengFig2}, the cut-off spatial frequency in heterostructures is higher than that in free space.  This property will lead to a higher resolution when transmitting images through the metamaterials.  For simplicity, we assume the periodic pattern has the same spatial period in the $x$ and $y$ directions.  Substituting Eq.~(\ref{beta}) into Eq.~(\ref{H2}) and changing the integration into double summations since the spectrum is discrete.  We obtained the Talbot distance in the layered heterostructures as 
\beq
\label{zT}
Z_T = {a^2d\sin(\beta_0d)\over\pi{\cal Y}(k_1,k_2)} \ ,
\eeq
where $a$ is the spatial period of the pattern, and ${\cal Y}(k_1,k_2)$ is given by
\beq
\label{Ypp}
{\cal Y}(k_1,k_2) = \left({\partial^2Y\over\partial k_x^2}\right)_{k_\bot=0} = \left({\partial^2Y\over\partial k_y^2}\right)_{k_\bot=0} = A + QP_1 + PQ_1 \ .
\eeq
The coefficients read,
\begin{eqnarray}
\label{QP}
A &=& \left({d_1\over k_1} + {d_2\over k_2}\right) \sin(k_1d_1+k_2d_2),  \cr  \noalign{\medskip}
Q &=& -2\sin(k_1d_1)\sin(k_2d_2),  \cr \noalign{\medskip}
P &=& {1\over4} \left(F+{1\over F} -2\right), \hskip.3in  F\equiv {k_1\epsilon_2\over k_2\epsilon_1},  \cr
P_1 &=& {1\over4k_1k_2} \left( {\mu_1^2+\mu_2^2\over\mu_1\mu_2} - 
{\epsilon_1^2+\epsilon_2^2\over\epsilon_1\epsilon_2} \right),  \cr \noalign{\medskip}
Q_1 &=& {2d_1\over k_1}\cos(k_1d_1)\sin(k_2d_2) + {2d_2\over k_2}\cos(k_2d_2)\sin(k_1d_1),
\end{eqnarray}
where $k_i^2=(\omega/c)^2\epsilon_i\mu_i$, $i=1,2$.

To confirm our theoretical predictions, numerical simulations of fractional Talbot images are demonstrated in Fig.~\ref{FengFig3} for the PNI stack, in Fig.~\ref{FengFig4} for the MD stack, and in Fig.~\ref{FengFig5} for the DD stack.  In those figures, the left plot is the original 2D square array while the right plot is the corresponding image at the fractional Talbot distance.  The numerical results were obtained from Eq.~(\ref{H2}) where the $\beta$ was solved numerically from Eq.~(\ref{disp}), not from Eq.~(\ref{beta}).  Unlike in free space where the paraxial approximation is required to observe Talbot images, upon carrying out the integration in Eq.~(\ref{H2}) no paraxial approximation was used.  Equation~(\ref{beta}) was used only for providing insight on how to effectively choose simulation parameters and for comparing the exact and paraxial results.  Further, in our simulations the sizes of the squares in the patterns are less than the illumination wavelength.

The interesting connection between the fractional Talbot effects and number theory can be used to explain the image patterns in Figs.~\ref{FengFig3}--\ref{FengFig5}.  The fractional Talbot image can be represented as a finite sum of spatially shifted subsidiary waves of the source field\cite{Berry,Banaszek}:
\beq
\label{sum}
E\left(x,y,{p\over q}Z_T\right) = \sum_{s=0}^{l-1} b_s E\left(x-{sa\over l}, y-{sa\over l}, 0\right),
\eeq
where $l=q/2$ if $q$ is a multiple of 4, and $l=q$ otherwise, and
\beq
\label{bs}
b_s = \sum_{n=0}^{l-1} \exp\left(-i2\pi{p\over q}n^2 -i2\pi n{s\over l}\right).
\eeq
For example in Fig.~\ref{FengFig3}a where $p/q=1/3$, the image is composed of three subsidiary waves $E(x,y)$, $E(x-a/3,y-a/3)$, and $E(x-2a/3,y-2a/3)$.  Thus, the spatial frequency is three times the original frequency.  In Fig.~\ref{FengFig3}b where $p/q=1/6$, the non-zero components are $b_1$, $b_3$, and $b_5$, so the image is also composed of three subsidiary waves, $E(x-a/6,y-a/6)$, $E(x-a/2,y-a/2)$, and $E(x-5a/6,y-5a/6)$.  Hence, the image in Fig.~\ref{FengFig3}b has the same spatial frequency as that in Fig.~\ref{FengFig3}a, but is spatially shifted by a half period.  In realistic nanoplasmonic structures, material loss cannot be avoided.  The effect of material loss can be mitigated by introducing gain inside the medium\cite{Ramak}.  Typically, the damping factor of metals is much smaller than the plasma frequency\cite{Scalora}, $\gamma\sim0.01\omega_e$.  As a demonstration, in Fig.~\ref{FengFig4} we compare the fractional Talbot images in the MD stack when $\gamma=0.0$\,fs$^{-1}$, $\gamma=0.1$\,fs$^{-1}$ (loss), and $\gamma=-0.1$\,fs$^{-1}$ (gain).  In the presence of loss or gain, the Talbot distance is a complex number, thus we redefine the Talbot distance as $\widetilde z_T=|z_T|^2/\Re(z_T)$.  The Talbot image is slightly blurred when there is a loss (Fig.~\ref{FengFig4}b), and is slightly sharper when there is a gain (Fig.~\ref{FengFig4}c).  It is well known that at a half Talbot distance the Talbot image has a reversed contrast compared to the original pattern.  This phenomenon is illustrated in  Fig.~\ref{FengFig5} in the DD structure.  For practical applications, we found when the error of layer thickness is less than 1\%, the Talbot images can still be observed.  Any interlayer width variation is expected to have a minimal effect since $Z_T\gg d$.  Since femtosecond lasers are widely used in material fabrication, for the illumination wavelength used in the DD and MD stacks, the Talbot images are sustainable with a wavelength variation of 7\% for the DD stack and 3\% for the MD stack.  This approximately corresponds to a 100 fs laser pulses.

In conclusion we have demonstrated subwavelength-scale fractional Talbot effects in layered heterostructures of metamaterials without using the paraxial approximation.  A general expression of the Talbot distance in such structures was obtained.  This expression can be used in potential applications involving Talbot effects and multilayer structures of metamaterials.  The fractional Talbot effect can be explored in material fabrication to increase the spatial frequencies of the periodic patterns.  This work is supported by NAVAIR's In-House Laboratory Independent Research (ILIR) program sponsored by the Office of Naval Research.

\eject
\newpage

\bfg[h]
\centerline{\scalebox{.6}{\includegraphics{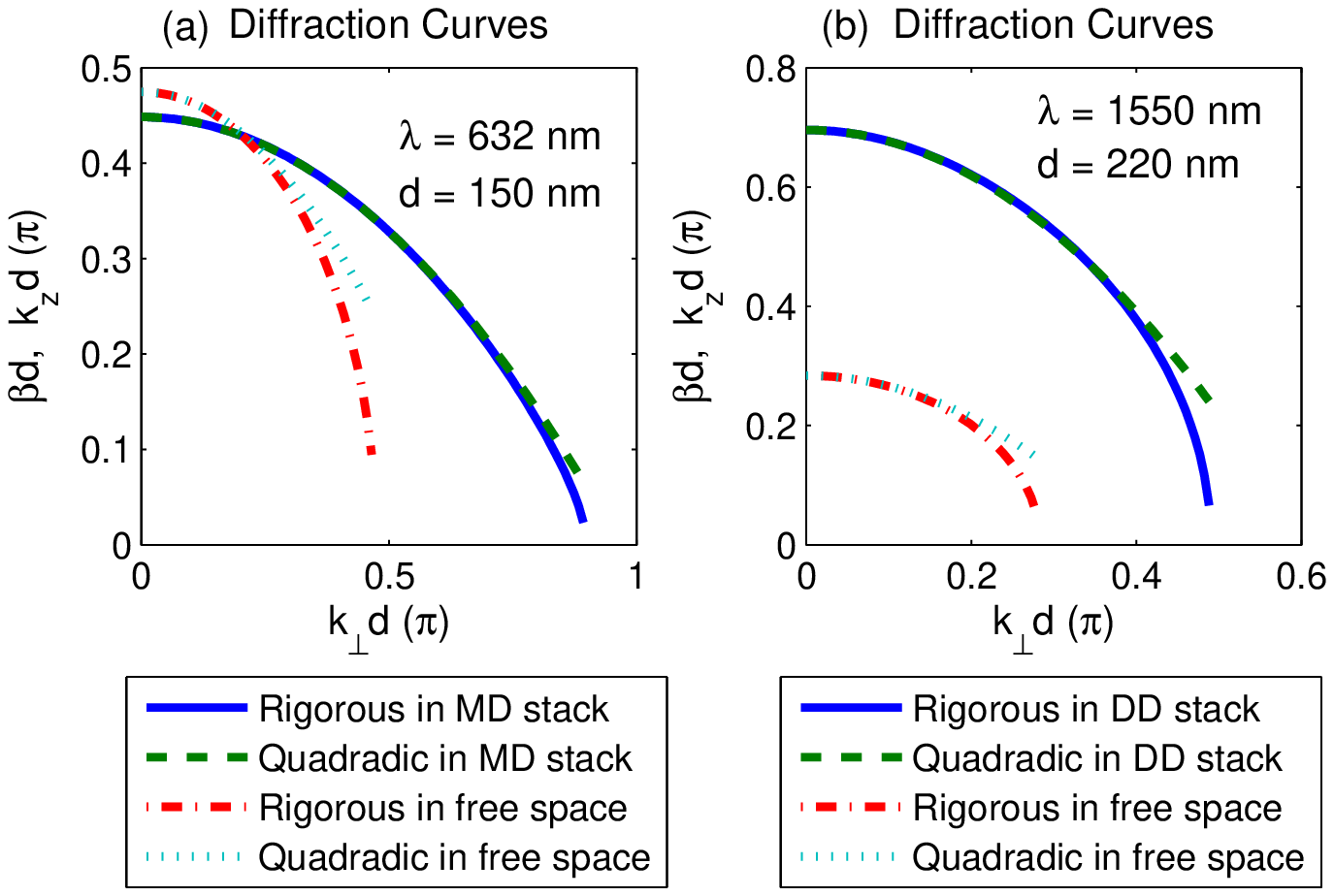}}}
\vskip-.2in
\caption{(Color online) Diffraction curves in the metallodielectric (a) and double-dielectric (b) stacks.  The blue solid curve is exact from Eq.~(\ref{disp}).  The green dashed curve is the quadratic approximation Eq.~(\ref{beta}).  The parameters in (a): $\lambda=632$\,nm, $\epsilon_1=2.66, \mu_1=\mu_2=1, \gamma_e=0.0\mbox{\,fs}^{-1}$, and $\omega_e=9.8\mbox{\,fs}^{-1}$ which gives $\epsilon_2=-9.797$ from the Drude model, $d_1=120$\,nm, and $d_2=30$\,nm; in (b): $\lambda=1550$\,nm, $\epsilon_1=1, \mu_1=\mu_2=1, \epsilon_2=7.6, d_1=60$\,nm and $d_2=160$\,nm.  As a comparison, the free-space diffraction is shown in the red dash-dotted curve (exact) and the cyan dotted curve (quadratic).  \label{FengFig1}}
\efg

\bfg[h]
\centerline{\scalebox{.6}{\includegraphics{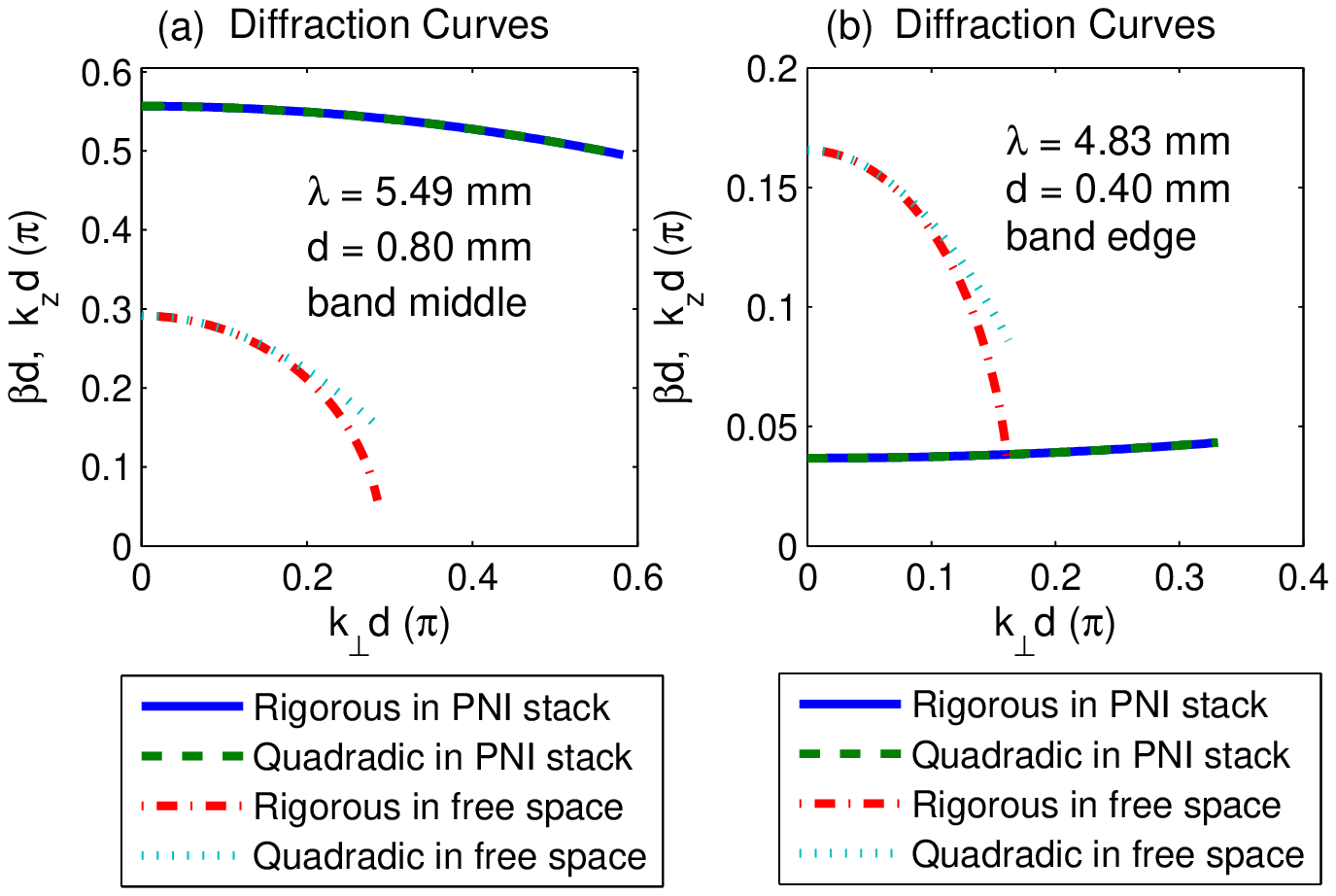}}}
\vskip-.2in
\caption{(Color online) Diffraction curves in the multilayer positive and negative index materials.  $\epsilon_1=2.66, \mu_1=1, \omega_e=780$\,GHz, $\omega_m=0.8\omega_e$, and $\gamma_e=\gamma_m=0.0\mbox{\,fs}^{-1}$.  (a) near the middle of the band, $\lambda=5.49$\,mm, $\epsilon_2=-4.165, \mu_2=-2.306, d_1=0.2$\,mm, $d_2=0.6$\,mm.  (b) near the band edge, $\lambda=4.83$\,mm, $\epsilon_2=-3, \mu_2=-1.56, d_1=d_2=0.2$\,mm.  The coincidence of the blue solid curve (exact) with the green dashed curve (quadratic) indicates the diffraction is quadratic in the PNI stack.  As a comparison, the free-space diffraction is shown in the red dash-dotted curve (exact) and the cyan dotted curve (quadratic).  \label{FengFig2}}
\efg

\bfg[h]
\centerline{\scalebox{.45}{\includegraphics{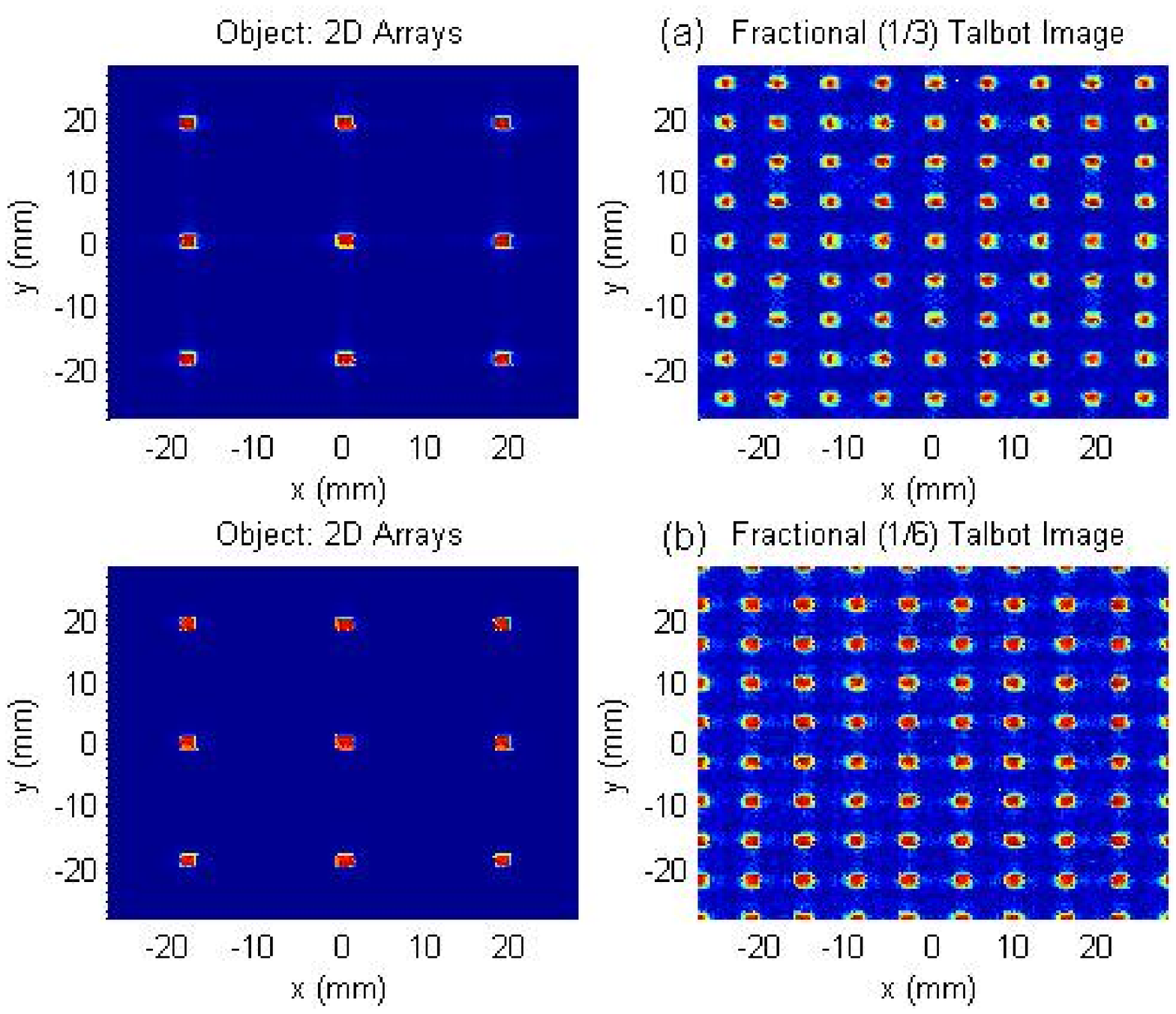}}}
\caption{(Color online) Fractional Talbot images in the PNI stack.  The size of the square is 2 mm.  (a) $z=Z_T/3$, $Z_T=1.2$\,m, all parameters are the same as those in Fig.~\ref{FengFig2}a (the middle of the band).  (b) $z=Z_T/6$, $Z_T=7.3$\,m, all parameters are the same as those in Fig.~\ref{FengFig2}b (the edge of the band).  Due to the phase compensation of the negative index medium, the Talbot distance is long.  \label{FengFig3}}
\efg

\bfg[h]
\centerline{\scalebox{.45}{\includegraphics{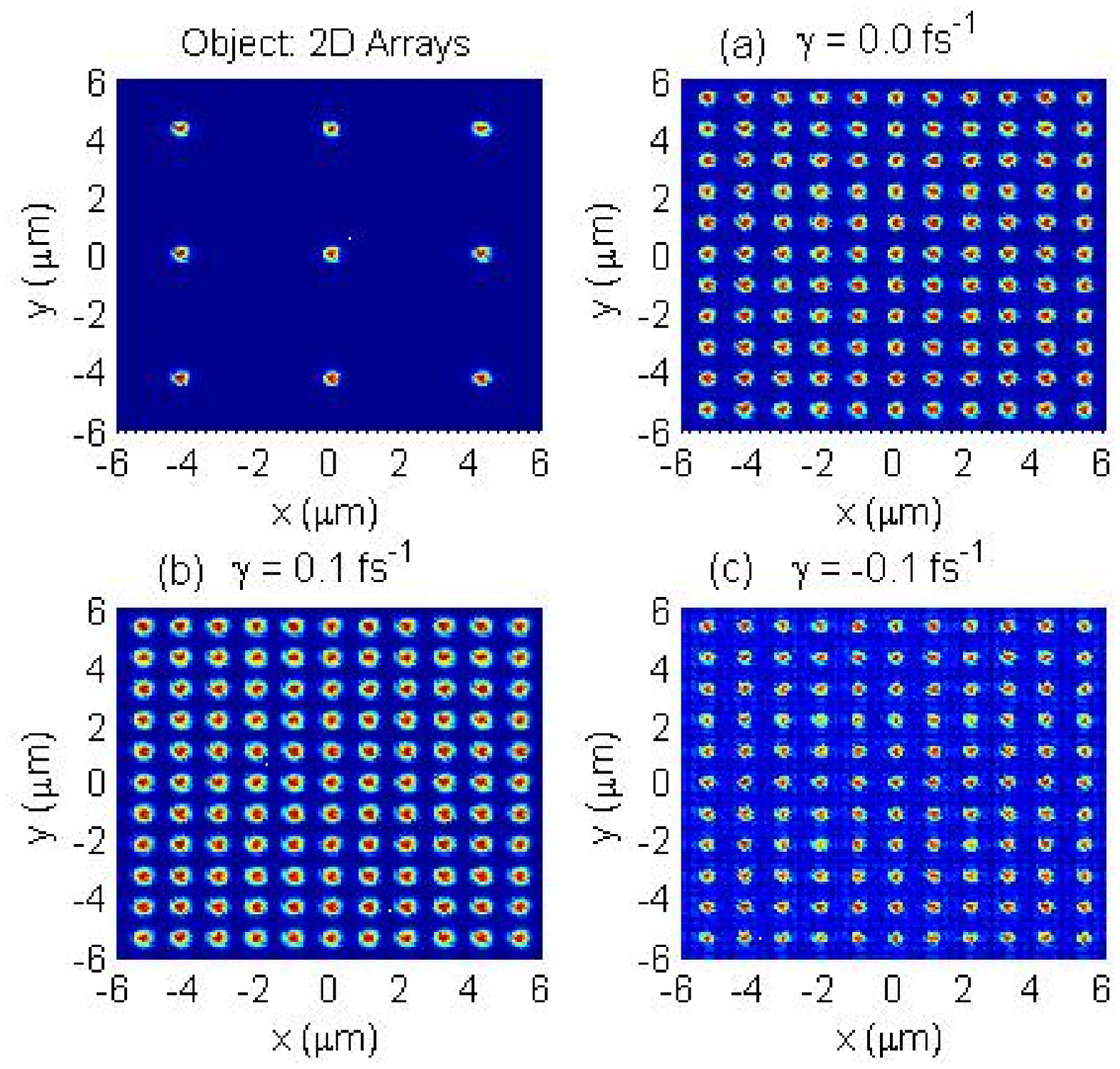}}}
\caption{(Color online) Fractional Talbot image in the MD stack in the presence of loss or gain.  The size of the square is $0.4\,\mu$m, $z=Z_T/8$, and $\omega_e=9.8$\,fs$^{-1}$.  (a) $\gamma=0.0$\,fs$^{-1}$, $Z_T=125.1\,\mu$m; (b) $\gamma=0.1$\,fs$^{-1}$ (loss), $\widetilde Z_T=125.3\,\mu$m;  (c) $\gamma=-0.1$\,fs$^{-1}$ (gain), $\widetilde Z_T=125.3\,\mu$m.  All other parameters are the same as those in Fig.~\ref{FengFig1}a.  \label{FengFig4}}
\efg

\bfg[h]
\centerline{\scalebox{.45}{\includegraphics{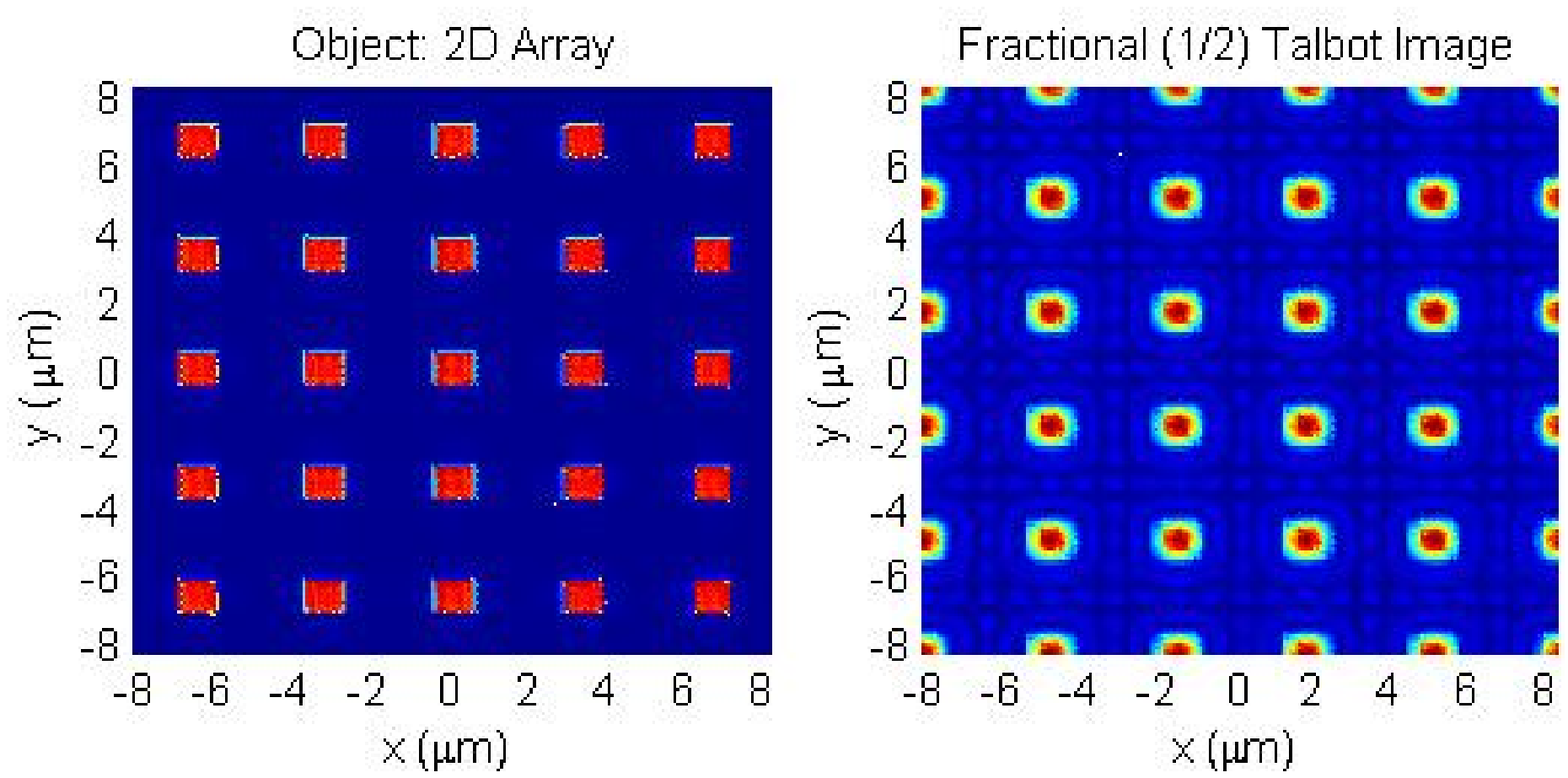}}}
\vskip-.3in
\caption{(Color online) Fractional Talbot image in the DD stack.  At a half Talbot distance the image has undergone a contrast reversal compared to the original pattern.  The size of the square is $1\,\mu$m.  $z=Z_T/2$, $Z_T=12.9\,\mu$m, all other parameters are the same as those in Fig.~\ref{FengFig1}b.  \label{FengFig5}}
\efg


\begin{thebibliography}{99}
\bibitem{Note}  We have confirmed numerically that there is a negligible difference when taking the Talbot image at the distance $z=Z_T$ and $z=Z_T\pm d$.
\end{thebibliography}
\end{document}